\documentclass[12pt]{article}
\usepackage[dvips]{color}
\usepackage{epsfig}
\usepackage{amsmath}
\usepackage{graphicx}

\begin{document}
\title{Strong Gravitational Lensing with Gauss-Bonnet correction}
\author{{J. Sadeghi $^{a}$\thanks{Email: pouriya@ipm.ir},\hspace{1mm} }
  and \hspace{1mm}H. Vaez $^{a}$
\thanks{Email: h.vaez@umz.ac.ir}\hspace{1mm} \\
{$^{a}$ \emph{Physics Department, Mazandaran University},}\\{
\emph{P.O.Box 47416-95447, Babolsar, Iran}}\\  } \maketitle
\begin{abstract}
In this paper we investigate the strong gravitational lensing in a
five dimensional background with Gauss-Bonnet gravity, so that in
4-dimensions the Gauss-Bonnet correction disappears. By considering
the logarithmic term for deflection angle, we obtain the deflection
angle $\hat\alpha$ and corresponding parameters $\bar{a}$ and
$\bar{b}$. Finally, we estimate some properties of relativistic
images such as $\theta_{\infty}$, $s$ and $r_m$. \noindent
 \\
\end{abstract}
\section{Introduction}
The deviation of the light rays in the gravitational fields is
referred to gravitational lensing. The gravitational lensing (GL) in
the weak limit has been used to test the General Relativity since
its beginning \cite{Einstein,schneider}. But, this theory in the
weak limit was not able to describe the high bending and looping of
the light rays. Hence, scientist community stated  this phenomenon
in the strong filed regime. In the strong field limit, the light
rays pass very close to black hole and one set of infinitive
relativistic "ghost" images would be produce on each side of black
hole. These images are produced due to the light rays wind one or
several times around the black hole before reaching to observer. At
first, this phenomenon was proposed by Darwin \cite{Darwin}. Several
studies of null geodesics in the strong gravitational field have
been done in the past years \cite{Bardeen}-\cite{Falcke}. In 2000,
Virbhadra and Ellis showed that a source of light behind a
schwarzschild black hole would product an infinitive series of
images on each side of the massive object \cite{Virbhadra1}. Theses
relativistic images are formed when the light rays travel very close
to the black hole horizon, wind several times around the black hole
before appearing at observer. By  an alternative method, Frittelli
et al. obtained an exact lens equation, integral expression for
deflection and compared their results with Virbhadra et al
\cite{Frittelli}. A new technic was proposed by Bozza et al. to find
the position of the relativistic images and their magnification
\cite{bozza2}. They used the first two terms of approximation to
study schwarzschild black hole lensing. This method was applied to
other works such as Eiroa, Romero and Torres studied  a
Reissner-Nordstrom black hole lensing\cite{Eiroa}; Petters
calculated relativistic effects on microlensing events
\cite{petters}. Afterward, the generalization of Bozza's method for
spherically symetric metric was developed in \cite{bozza1}. Bozza
compared the image patterns for several interesting backgrounds and
showed that by the separation of the first two relativistic images,
we can distinguish two different collapsed objects. Further studies
were developed  for other black holes and metrics
\cite{bozza5}-\cite{saadat}.\\
The gravitational lenses are important tools for probing the
universe. In Refs. \cite{NarasimhaChitre,Narasimha} Narasimha and
Chitre predicted that the gravitational lensing of dark matter can
give the useful data about the position of the dark matter in the
universe . Also, the gravitational lens are used to detect the
exotic objects in the universe,
 such as cosmic strings \cite{hogan}-\cite{Vilenkin2}.\\
 On the other hand, the gravitational theories in higher dimensions have  attracted
 considerable attention. One of these higher dimension gravities is
 the supersymmetric  string theory. Einstein-Gauss-Bonnet ($EGB$) theory, which emerges as the low-energy limit of this
 theory, can be considered as an effective model  of gravity in
 higher dimensions. This theory yields a correction to
 Einstein-Hilbert action. The Gauss-Bonnet term involves up to
 second- order derivatives of the metric with the same degrees of
 freedom as the Einstein theory \cite{Zwiebach,zumino}. The variation of EGB action has different
 solutions and the spherically symmetric solution  in the presence of Gauss-Bonnet gravity was obtained by Boulwar
 and Deser \cite{Boulware} and charged black hole one is found by Wiltshire \cite{wiltshire}. The properties of the Gauss-Bonnet black
 holes have been studied in Refs.
 \cite{Rong1}-\cite{Ishwaree}.\\
 In this paper we study the strong
 gravitational lensing and obtain the logarithmic deflection angle and corresponding coefficients.
 In the final we investigated some properties of relativistic
 images.
  \\ The paper is organized as follows: In Section 2, we briefly present the
  Einstein-Gauss-Bonnet gravity. Section 3 is devoted to
  investigate  the strong  gravitational lensing in the presence of Gauss-Bonnet term.
  We consider the logarithmic term which was proposed by Bozza, and
  obtain its parameters $\bar{a}$ and $\bar{b}$. In section 4, some
  properties of relativistic images will be studied. Finally, in the
  last section we present summery.
\section{Einstein-Gauss-Bonnet gravity }
The action of Einstein-Gauss-Bonnet gravity in five dimensional is
given by \cite{Boulware},
\begin{equation}\label{action}
I=-\frac{1}{16\pi G_5}\int{d^5 x\sqrt{-g}\,\left(R+\frac{\alpha}{2}
L_{GB}\right)},
\end{equation}
where $R$ and $\alpha$ are Ricci scalar and Gauss-Bonnet constant
respectively. $G_5$ is five-dimensional Newton's constant and
$L_{GB}$ is the Gauss-Bonnet term as follows,
\begin{equation}\label{GB}
L_{GB}=R^2-4R_{ab}R^{ab}+R_{abcd}R^{abcd},
\end{equation}
here $R_{ab}$ and $R_{abcd}$ are Ricci tensor and Riemann tensor
respectively. Note that the indexes run over the components of five
dimensional space. The exact and spherically metric solution of the
above action have been founded by Boulware and Deser
\cite{Boulware},
\begin{equation}\label{metric1}
ds^2=-f(r)dt^2+f(r)^{-1}dr^2+c(r)d\Omega_3^2\,\,,
\end{equation}
where
\begin{eqnarray}\label{f(r)}
f(r)=1+\frac{r^2}{2\alpha}\left(1-\sqrt{1+\frac{8\alpha
M}{r^4}}\right),\quad\quad c(r)=r^2.
\end{eqnarray}
\begin{figure}
\centerline{\epsffile{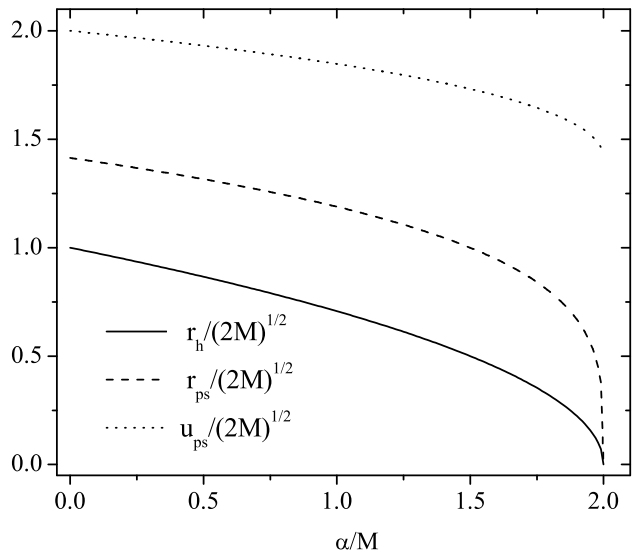}} \caption{The figure shows the
variation  of horizon, photon sphere radius and minimum impact
parameter with respect to $\alpha$.
  }\end{figure}
Here $M$ is related to $ADM$ mass and note that we set $G = c = 1$.
  For simplicity, we introduce the
dimensionless quantities as $a=\frac{\alpha}{M}$ and
$x=\frac{r}{\sqrt{2M}}$. So, we have,
\begin{eqnarray}\label{f(x)}
f(x)=1+\frac{x^2}{a}\left(1-\sqrt{1+\frac{2a}{x^4}}\right).
\end{eqnarray}
When $a$ tends to zero the warp factor of the Myers-Perry metric
 is obtained \cite{Eiroa2}. The solution of $f(x)=0$,
$x_h=\frac{1}{2}\sqrt{4-2a}$ is the horizon radius of the black
hole. The variation of the horizon is plotted with respect to $a/M$
in figure 1.

\epsfxsize=12cm \epsfysize=10cm

\section{Lens equation, Deflection angle with Gauss-Bonnet correction }
The lens equation for a source of light and an observer situated at
large distances from a lens(deflector) is given by \cite{Bozza2008},
\begin{equation}\label{horizon1}
D_{os}\tan\beta=\frac{D_{ol}\sin\theta-D_{ls}\sin(\hat\alpha-\theta)}{\cos(\hat\alpha-\theta)}.
\end{equation}
Where, $D_{ls}$ and $D_{os}$ stand for the lens-source and
observer-source diameter distance, respectively. The angular
positions of source and images with respect to the optical axis (the
line joining the observer and center of the lens) are represented by
$\beta$ and $\theta$. The deflection of the light rays denotes by
$\hat{\alpha}$ which can be positive, $\hat{\alpha}>0$ (bending
toward the lens) or be negative, $\hat{\alpha}<0$ (bending away from
the lens). In the next section, we will obtain the deflection angle.
The particular distance from the center of the lens to the null
geodesic at the source position is called impact parameter, which is
given by following expression,
\begin{eqnarray}\label{J}
u=D_{ol}\,\sin\theta.
\end{eqnarray}
We can find the angular positions of images by the intersection of
two functions $\tan\theta-\tan\beta $ and $
\frac{D_{ls}}{D_{os}}(\tan\theta+\tan(\hat{\alpha}-\theta))$ vs
$\theta$ for the same side and vs $-\theta$ for opposite side. In
addition to the primary and secondary image positions (due to the
weak limit), there is a sequence of intersections  that show the
angular positions of the relativistic images. These points are very
close to each other, so they are not distinguishable. For this
reason, we call them relativistic images. These images are due to
the bending of light rays more than $3\pi/2$.
\\Now, we are going to investigate the deflection angle in the presence
of Gauss-Bonnet correction gravity. By using the null geodesic
equation for the following standard background metric,
\begin{equation}\label{metric3}
ds^2=-\mathcal{A}(r)dt^2+\mathcal{A}^{-1}(r)dr^2+\mathcal{C}(r)\,d\phi^2+\mathcal{D}(r)d\psi^2,
\end{equation}
one can find the  following equations,
\begin{eqnarray}\label{constMo}
&&\dot{t}=\frac{E}{\mathcal{A}(r)},\nonumber\\
&&\dot{\phi}=\frac{L_\phi}{\mathcal{C}(r)},\nonumber\\
&&\dot{\psi}=\frac{L_\psi}{\mathcal{D}(r)},
\end{eqnarray}
\begin{equation}\label{rGeo}
(\dot{r})^2=\frac{1}{\mathcal{B}(r)}\left[\frac{\mathcal{D}(r)E-\mathcal{A}(r)L^2_\psi}{\mathcal{A}(r)\mathcal{D}(r)}-\frac{L^2_\phi}{\mathcal{C}(r)}\right].
\end{equation}
where $E$ is the energy of photon and $L_\phi$ and $L_\psi$ are
angular momentums in $\phi$ and $\psi$ directions. Here a dot
denotes derivation with respect to affine parameter. If we consider
the $\theta$ component of geodesic equations in the equatorial plane
$(\theta=\pi/2$), we have
\begin{eqnarray}\label{s14}
\dot{\phi}\left[\mathcal{D}(r)\dot{\psi}\right]= \dot{\phi}L_\psi=0.
\end{eqnarray}
Here,  if we consider $\dot{\phi}=0$, the deflection angle of light
ray becomes zero and this is illegal, therefor we set $L_\psi=0$.
For a light ray coming from infinity the deflection angle in the
directions $\phi$  is given by \cite{Virb},
\begin{eqnarray}\label{alpha2}
\hat{\alpha_\phi}=I_\phi(x_0)-\pi,\nonumber\\
\end{eqnarray}
where
\begin{eqnarray}\label{alpha}
I(x_0)=2\int^\infty_{x_0}\left[\frac{{\mathcal{C}}(x)}{{\mathcal{C}}(x_0)}{\mathcal{A}}(x_0)-\mathcal{A}(x)\right]^
{-\frac{1}{2}}\frac{dx}{x},
\end{eqnarray}
where $x_0$ is the closet approach distance for the light ray when
it passes  near the lens. The impact parameter for the closet
approach is expressed by,
\begin{eqnarray}\label{Impact}
u(x_0)=\sqrt{\frac{\mathcal{C}(x_0)}{\mathcal{A}(x_0)}}=x\sqrt{\frac{1}{1+\frac{x^2(1-\sqrt{1+\frac{2a}{x^4}})}{a}}}.
\end{eqnarray}
The above relation is obtained from the null geodesic equation
(\ref{rGeo}) with setting $dr/d\phi=0.$ By using (\ref{J}) and
(\ref{Impact}) one can relate the image position to the closet
approach and this relation allows us to write the deflection angle
as a function of image position. The image position is plotted as a
function of the closest approach in figure 2. We see that the image
positions and distances between relativistic images reduce by
increasing the Gauss-Bonnet parameter. For the large values of $x_0$
the curves coincide for any value of Gauss-Bonnet parameter and this
means that primary and secondary image position remain unchanged.
\begin{figure}
\centerline{\epsffile{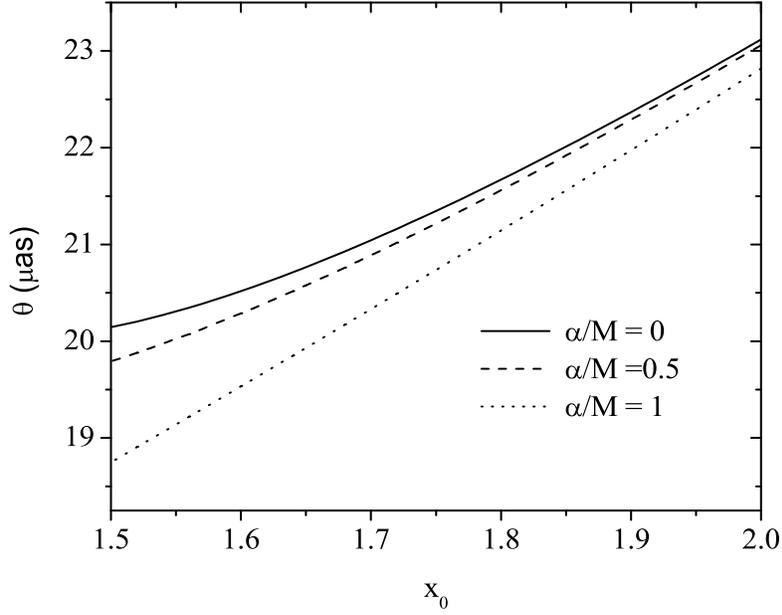}} \caption{The angular position
of images with respect to $x_0$ at $\alpha/M=0$, $\alpha/M=.5$ and
$\alpha/M= 1$. (Mass 4,31$\times10^6M_\odot$, the distance
$D_{ol}=8.5 Kpc$, and $\mu as\equiv$microarcseconds)}
\end{figure}
There is a minimum value for the closest approach that is called the
photon sphere radius and is a $r=$const null geodesic. The photon
sphere is the root of derivative of the impact parameter with
respect to $x_0$ which is given by,
\begin{eqnarray}\label{function}
x_{ps}=\,(4-2a)^{\frac{1}{4}}.
\end{eqnarray}
The dashed curve in the figure 1 shows the variation of  photon
sphere radius. It decreases with increasing the Gauss-Bonnet
parameter and tends to zero at $\alpha=2$. When $x_0$ asymptotically
approaches the photon sphere radius, the photon reveals around the
lens more times and the deflection angle diverges as $x_0$ tends to
photon sphere. We can rewrite the equation~(\ref{alpha}) as,
\begin{equation}\label{Iz}
I(x_0)=2\int^1_0F(z,x_0)\,dz,
\end{equation}
and
\begin{equation}\label{f(z)}
F(z,x_0)=\frac{1}{\sqrt{\mathcal{A}(x_0)-\mathcal{A}(x)\frac{\mathcal{C}(x_0)}{\mathcal{C}(x)}}},
\end{equation}
where $z=1-\frac{x_0}{x}$. The function  $F(z,x_0)$ diverges as $z$
approaches to zero. Therefore, we can split the integral~(\ref{Iz})
in two parts, the divergent part $I_D(x_0)$ and the regular one
$I_R(x_0)$, as follows \cite{bozza1}
\begin{equation}\label{Id}
I_D(x_0)=2\int^1_0F_0(z,x_0)\,dz,
\end{equation}
\begin{equation}\label{Ir}
I_R(x_0)=2\int^1_0\left[F(z,x_0)-F_0(z,x_0)\right]\,dz.
\end{equation}
Here we expand the argument of the square root in $F(z,x_0)$ up to
the second order in $z$
\begin{equation}\label{f0}
F_0(z,x_0)=\frac{1}{\sqrt{p(x_0)z+q(x_0)z^2}},
\end{equation}
where
\begin{eqnarray}\label{p}
&p(x_0)=\frac{x_0}{c(x_0)}\left[c^\prime(x_0)f(x_0)-c(x_0)f^\prime(x_0)\right]\nonumber\\
&=
-\frac{2\left(-x_0^4-2a+2x_0^2\sqrt{\frac{x_0^4+2a}{x_0^4}}\right)}{x_0^4+2a}
\end{eqnarray}
\begin{eqnarray}\label{q}
&q(x_0)=\frac{x_0^2}{2c(x_0)}\left[2c^\prime(x_0)c(x_0)f^\prime(x_0)
-2c^\prime(x_0)^2f(x_0)+f(x_0)c(x_0)c^{\prime\prime}(x_0)-
c^2(x_0)f^{\prime\prime}(x_0)\right] \nonumber\\
&=\frac{-x_0^8-4x_0^4a-4a^2+6x_0^6\sqrt{\frac{x_0^4+2a}{x_0^4}}+4x0^2\sqrt{\frac{x_0^4+2a}{x_0^4}}a}{(x_0^4+2a)^2}.
\end{eqnarray}
As $a$ goes to zero, $p$ and $q$ tend to five dimensional
schwarzschild ones, $p=-\frac{4}{x0^2}+2$ and $q=\frac{6}{x0^2}-1$.
For $x_0>x_{ps}$, $p(x_0)$ is nonzero and the leading order of the
divergence in $F_0$ is $z^{-1/2}$, which have a finite result. As
$x_0 \longrightarrow x_{ps}$, $p(x_0)$ approaches zero and the
divergence is of order $z^{-1}$, that makes the integral divergent
logarithmically . Therefor, the deflection angle can be approximated
in the following form \cite{bozza1}
\begin{equation}\label{deflection}
\hat{\alpha}=-{\bar{a}}\,\log\left(\frac{u}{u_{ps}}-1\right)+{\bar{b}}+O(u-u_{ps}),
\end{equation}
\epsfxsize=12cm \epsfysize=10cm
\begin{figure}
\centerline{\epsffile{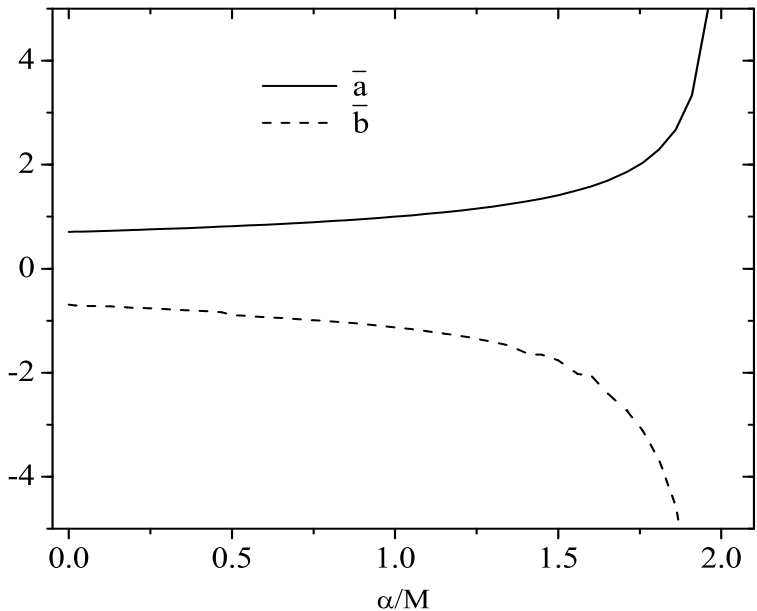}} \caption{The coefficients $\bar{a}$
and $\bar{b}$ as  functions of the Gauss-Bonnet parameter.}
\end{figure}
where
\begin{eqnarray}\label{a}
&\bar{a}=\frac{1}{\sqrt{q(x_{ps})}}\,\approx\frac{\sqrt{2}}{2}+0.128\,a+0.161\, a^2\nonumber\\
&\bar{b}=-\pi+b_R+\bar{a}\,\log\frac{x_{ps}^2\left[\mathcal{C}^{\prime\prime}(x_{ps})\mathcal{F}(x_{ps})-
\mathcal{C}(x_{ps})\mathcal{F}^{\prime\prime}(x_{ps})\right]}{u_{ps}\sqrt{\mathcal{F}^3(x_{ps})\mathcal{C}(x_{ps})}}
\approx0.6902+0.154\,a+0.373\,a^2\,,\nonumber\\
&b_R=I_R(x_{ps}),\,\,\,\,\,\,u_{ps}=\sqrt{\frac{\mathcal{C}(x_{ps})}{\mathcal{F}(x_{ps})}}\,.
\end{eqnarray}
When $a$ tends to zero, we have $a=\frac{\sqrt{2}}{2}$ and
$b=0.6902$, that these values belong to Myers-Perry metric
\cite{Eiroa2}. Using ~(\ref{deflection}) and~(\ref{a}), we can
investigate the properties of strong gravitational lensing in the
presence of Gauss- Bonnet correction. The variations of the $u_{ps}$
is shown in figure 1. Also, coefficients $\bar{a}$, $\bar{b}$, and
the deflection angle $\hat{\alpha}$ have been plotted with respect
to
 the Gauss- Bonnet correction in
 figures 3-4. We see that by increasing $\alpha$, the deflection angle $\hat{\alpha}$ and $\bar{a}$  increase and $\bar{b}$ decreases. The deflection angle becomes
 diverge as $\alpha\longrightarrow2$.  \\
 \begin{figure}
\centerline{\epsffile{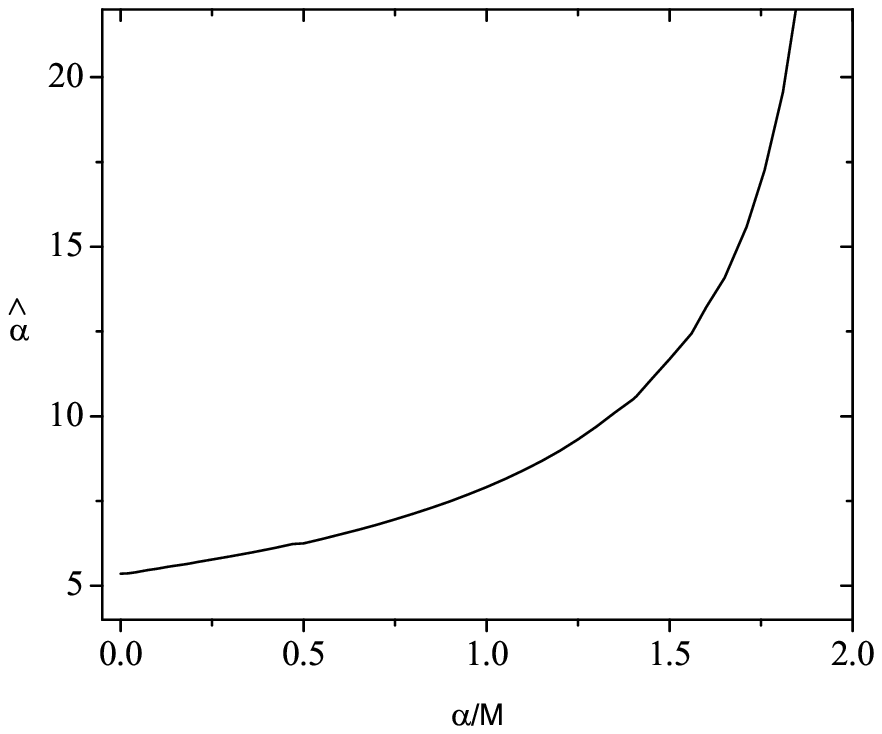}} \caption{Deflection angle in
presence of Gauss-Bonnet term at $x_0=1.01x_{ps}$.
  }
\end{figure}
\begin{figure}
\centerline{\epsffile{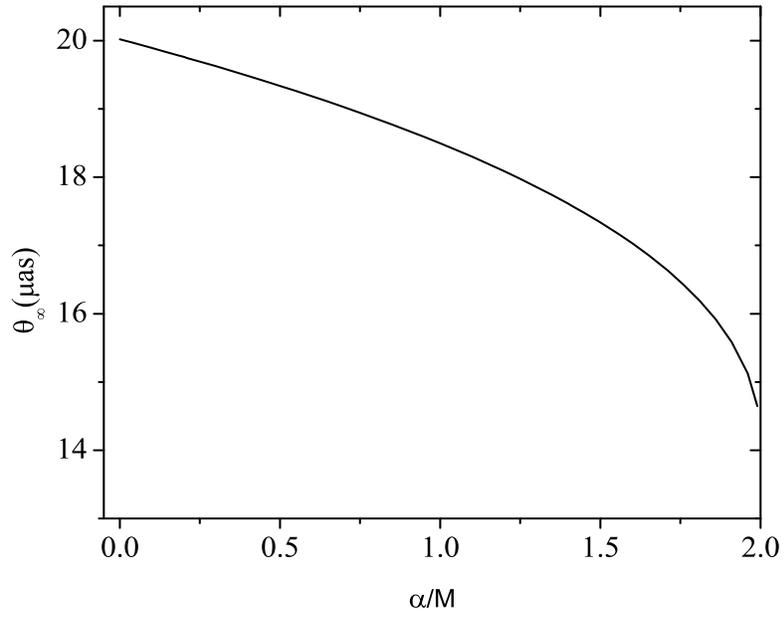}} \caption{The variation of
compacted images position as a function of Gauss-Bonnet parameter.
  }
\end{figure}
\begin{figure}
\centerline{\epsffile{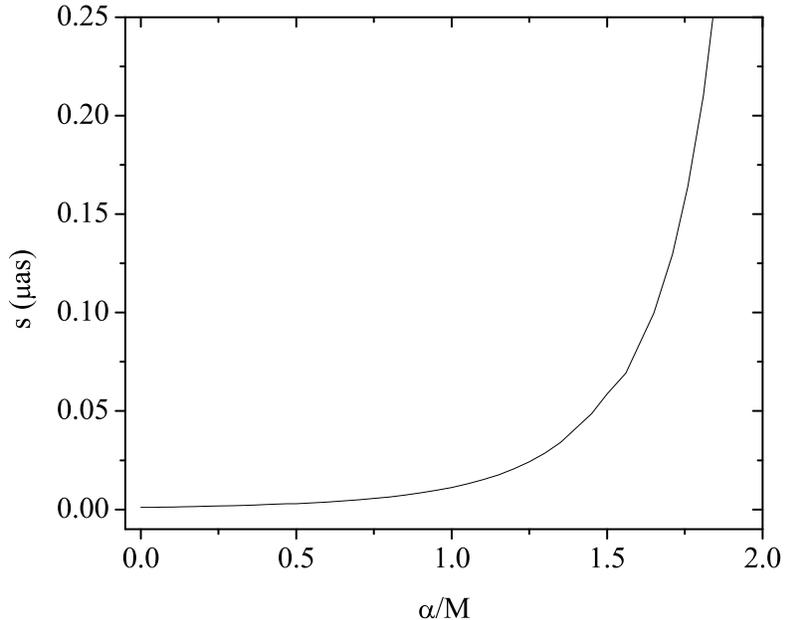}} \caption{The variation of angular
separation $s$ with respect to $\alpha$.
  }
\end{figure}
\begin{figure}
\centerline{\epsffile{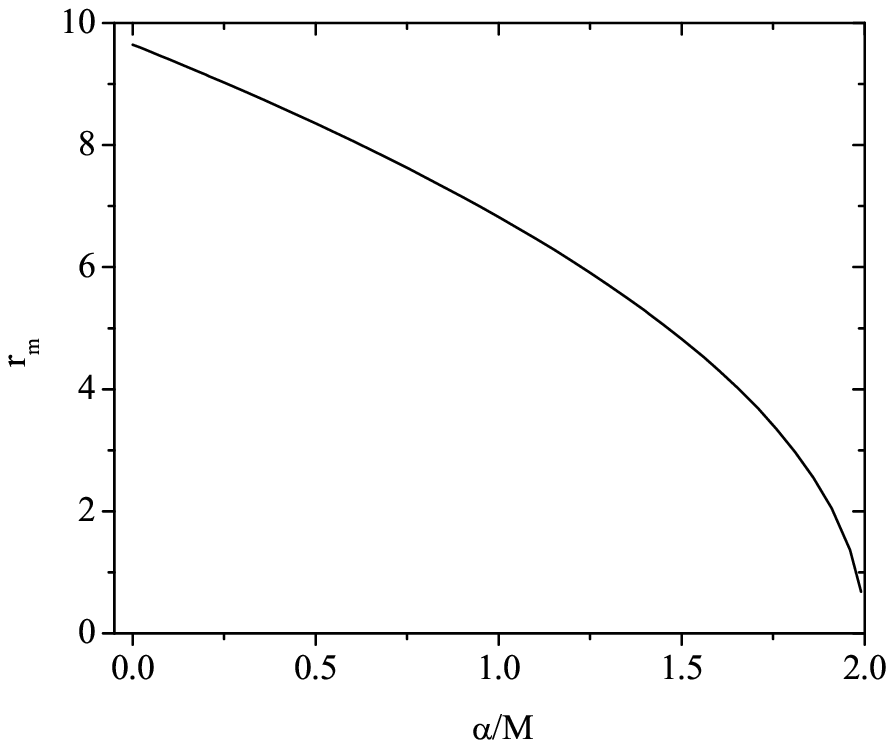}} \caption{The relative magnification
$r_m$ versus  $\alpha$.
  }
\end{figure}
\section{Relativistic images properties}
In the previous section,  we investigated the strong gravitational
lensing by using a simple and reliable logarithmic formula for
deflection angle that was obtained by Bozza et al. and obtained
corresponding parameters $\bar{a}$ and $\bar{b}$. Now we study some
properties of relativistic images in the presence of Gauss-Bonnet
gravity. When a source, lens, and observer are highly aligned, we
can write the lens equation in strong gravitational lensing,  as
following \cite{bozza1}
\begin{equation}\label{lensEQ}
\beta=\theta-\frac{D_{ls}}{D_{os}}\Delta\alpha_n,
\end{equation}
where $\Delta\alpha_n=\alpha-2n\pi$ is the offset of deflection
angle in which all the loops are subtracted, and the integer $n$
indicates the $n$-th image. The image position $\theta_n$ and the
image magnification $\mu_n$ can be approximated as obtained in Ref
\cite{bozza2},
\begin{equation}\label{theta}
\theta_n=\theta^0_n+\frac{u_{ps}(\beta-\theta_n^0)e^{\frac{\bar{b}-2n\pi}{\bar{a}}}D_{os}}{\bar{a}
D_{ls}D_{ol}},
\end{equation}
\begin{equation}\label{magnification}
\mu_n=\frac{u_{ps}^2(1+e^{\frac{\bar{b}-2n\pi}{\bar{a}}})e^{\frac{\bar{b}-2n\pi}{\bar{a}}}D_{os}}{\bar{a}\beta
D_{ls}D_{ol}^2},
\end{equation}
where
\begin{equation}\label{magnification}
\theta_m^0 = \theta_{ps}( 1 + e^{(\bar{b}-m\pi)/\bar{a}})  ,
\end{equation}
$\theta_n^0$ is the angular position of $\alpha=2n\pi$. They
separate the outer most image $\theta_1$ from the others images
which are packed together at $\theta_{\infty}$. Therefore, the
separation between $\theta_1$ and $\theta_{\infty}$ and ratio of
their magnification can be considered by,
\begin{eqnarray}\label{sR}
&&s=\theta_1-\theta_{\infty}\nonumber\\
&&\mathcal{R}=\frac{\mu_1}{\sum^{\infty}_{n=2}\mu_n}.
\end{eqnarray}
The asymptotic position  of the set of images $\theta_{\infty}$ can
be obtained from the minimum of the impact parameter as,
\begin{eqnarray}\label{thetainfty}
\theta_{\infty}=\frac{u_{ps}}{D_{ol}}.
\end{eqnarray}
By considering equation (\ref{thetainfty}), we can approximate
equations (\ref{sR}) as,
\begin{eqnarray}\label{sR2}
&&s=\theta_{\infty}e^{\frac{\bar{b}}{\bar{a}}-\frac{2\pi}{\bar{a}}},\nonumber\\
&&\mathcal{R}=e^{\frac{2\pi}{\bar{a}}}.
\end{eqnarray}
Another property that can be defined for relativistic images is the
relative magnification of the outermost relativistic image with the
other ones. This is shown by $r_m$ which is related to $\mathcal{R}
$ as,
\begin{eqnarray}\label{rm}
r_m=2.5\, \log\mathcal{R}\,.
\end{eqnarray}
If we suppose a five dimensional black hole with mass $4.31\times
10^6 M_\odot$ ( Galaxy center mass) and the distance between the
observer and black hole is $D_{OL}=8.5\,kpc$ (The distance between
the sun and galaxy center) \cite{Gillessen}, we can study the effect
of the Gauss-Bonnet parameter on these quantities. Our results are
presented in figure
4-6 and Table 1.\\

\begin{table*}\label{s}
\begin{center}
\begin{tabular}{  c c  c| c c   c c c  cc| cc  c  c c}
  \hline
  \hline
   & $\alpha/M$     && &    $ {\theta_\infty}$     & &  $s$ & &  $r_m$ & && $ a$     &&      $b$   &       \\
  \hline
   &$0$   &&& 20.02002002&&0.001043418 && 9.647597725&& &0.707106781 &&-0.690292235&\\
  & $0.3$ &&& 19.6255146 && 0.001970453&& 8.894645615&& &0.766964988 &&-0.777741402&\\
   &$0.6$ &&& 19.18509246&& 0.00377615 && 8.071759368&& &0.845154255&& -0.928671426 &\\
  & $0.9$ &&&18.68212151&&0.008402205&&7.15484996&&  &0.953462589&&-1.06498697&\\
  & $1.2$ &&&18.08715205&&0.020603386&&6.10167655&&  &1.118033989&&-1.294291956&   \\
  & $1.5$ &&&17.33784593&&0.058672906&&4.823798862&&&1.414213562&&-1.761807496&\\
  & $1.8$ &&&16.1916094 &&0.210208553&&2.973589325&&&2.294157338&&-3.6829744&\\
 &  $1.95$&&&15.12420331&&0.480609627&&1.364376354&&&5.00123&&-10.96179596&\\
  \hline
\end{tabular}
\caption {Numerical estimations for the coefficients and observables
of strong gravitational lensing with Gauss-Bonnet correction . (Not
that the numerical values for $\theta_{\infty}$ and $s$ are of order
microarcsec).}
\end{center}
\end{table*}

\section{Summary}
The light rays can be deviated from a straight way in the
gravitational field as predicted by General Relativity in which this
deflection of light rays is known as gravitational lensing. In the
strong field limit, the deflection angle of the light rays passing
very close to the black hole, becomes so large  that, the light rays
wind several times around the black hole before appearing at the
observer. Therefore the observer would detect two infinite set of
faint relativistic images produced on each side of the black hole.
On  the other hand, the gravitational theories in higher dimensions
have been attracting
 considerable attention in recent decades. Einstein-Gauss-Bonnet theory that emerges as the
low-energy limit of supersymmetric  string theory, is one of the
candidates for higher dimension theory. We considered five
dimensional metric with Gauss-Bonnet correction and studied the
strong gravitational lensing and obtained the deflection angle and
corresponding parameters $\bar{a}$ and $\bar{b}$. We saw that by
increasing $\alpha$, the deflection angle $\hat{\alpha}$ and
$\bar{a}$ increase and $\bar{b}$ decreases. The deflection angle
became
 diverge as $\alpha\longrightarrow2$.

  Finally, we estimated some properties of relativistic images
which can be detected by astronomical instruments. Our results have
been presented in Figures 5-7. In figures 5 and 7, the variations of
$\theta_{\infty}$ and $r_m$, shown that the position of compacted
images and relative magnification reduce with increasing $\alpha$.
Also the angular separation is an increasing function and diverges,
as $\alpha$ tends to two (figure 6). Furthermore,  we saw that the
position of images reduces with $\alpha$ (see figure 2). But you
should note the decreasing rate of $\theta_{\infty}$ is more than
$\theta_1$, therefore $s$ will be increasing function.


\begin{thebibliography}{11}
\bibitem{Einstein}A. Einstein, \emph{Science}, \textbf{84},  506 (1936).
\bibitem{schneider}P. Schneider, J. Ehlers, and E. E. Falco "Gravitational
Lenses" \emph{Springer-Verlag, Berlin}, (1992).
\bibitem{Darwin}C. Darwin, \emph{Proc. of the Royal Soc. of London} \textbf{249}, 180 (1959).

\bibitem{Bardeen}J. M. Bardeen, "Black Holes", ed. C. DeWitt and B. S. deWitt, \emph{Gordon and  Breach}, \textbf{215} (1973).
\bibitem{Viergutz} S.U. Viergutz,\emph{ A and A} \textbf{272}, 355 (1993).
\bibitem{nemiroff}R. J. Nemiroff,\emph{ Amer. Jour. Phys.} 61,619
(1993).
\bibitem{Falcke}H. Falcke, F. Melia and E. Agol, \emph{APJ Letters} \textbf{528} L13 (1999).
\bibitem{Virbhadra1}K. S. Virbhadra and G. F. R. Ellis, \emph{Phys. Rev. D} \textbf{62}, 084003 (2000).


\bibitem{Frittelli}S. Frittelli, T. P. Kling and E. T. Newman, \emph{Phys. Rev. D}
\textbf{61},  064021 (2000).
\bibitem{bozza2}V. Bozza, S. Capozziello, G. lovane and G. Scarpetta, \emph{Gen. Rel. and Grav}. \textbf{33}, 1535 (2001).
\bibitem{Eiroa}E. F. Eiroa, G. E. Romero and D. F. Torres, \emph{Phys. Rev. D} \textbf{66}, 024010 (2002); E. F. Eiroa, \emph{Phys. Rev. D} \textbf{71},
083010 (2005); E. F. Eiroa, \emph{Phys. Rev. D} \textbf{73}, 043002
(2006).

\bibitem{Eiroa2}E. F. Eiroa, \emph{Phys.Rev. D} \textbf{71}, 083010
(2005).
\bibitem{petters}A. O. Petters, \emph{MNRAS} \textbf{338}, 457 (2003).
\bibitem{bozza1}V. Bozza, \emph{Phys. Rev. D} \textbf{66}, 103001 (2002).
\bibitem{bozza5}V. Bozza, F. De Luca, G. Scarpetta and M. Sereno,
\emph{Phys. Rev. D} \textbf{72}, 08300 (2005);
\bibitem{bozza3}V. Bozza, \emph{Phys. Rev. D} \textbf{67}, 103006 (2003).
\bibitem{Bozza2008}V. Bozza, \emph{Phys. Rev. D} \textbf{78}, 103005
 (2008).
\bibitem{Bhadra}A. Bhadra, \emph{Phys. Rev. D} \textbf{67}, 103009 (2003).


\bibitem{Whisker}R. Whisker, \emph{Phys. Rev. D} \textbf{71}, 064004 (2005).


\bibitem{Ghosh}T. Ghosh and S. Sengupta, \emph{Phys. Rev. D} \textbf{81}, 044013 (2010).
\bibitem{Aliev}A. N. Aliev and P. Talazan, \emph{Phys. Rev. D} \textbf{80}, 044023 (2009).

\bibitem{galin}Galin N. Gyulchev and Ivan Zh. Stefanov, \emph{Phys. Rev. D}
\bibitem{naked}K. S. Virbhadra and G. F. R. Ellis, \emph{Phys. Rev. D} \textbf{65}, 103004 (2002).
\bibitem{Virbhadra2}K. S. Virbhadra, \emph{Phys. Rev. D} \textbf{79}, 083004 (2009).
\bibitem{Virbhadra3}K. S. Virbhadra and C. R. Keeton, \emph{Phys. Rev. D} \textbf{77}, 124014
(2008).
\bibitem{Virb}K. S. Virbhadra, D. Narasimha and S. M. Chitre,
\emph{Astron. Astrophys}. \textbf{337}, 1-8 (1998). 87, 063005
(2013).
\bibitem{SKKBH}Y. Liu, S. Chen and J. Jing, \emph{Phys. Rev. D} \textbf{81}, 124017 (2010).
\bibitem{SKKGBH}S. Chen, Y. Liu and J. Jing, \emph{Phys. Rev. D} \textbf{83}, 124019
(2011).
\bibitem{CSKKBH}J. sadeghi, A. Banijamali and H. Vaez, \emph{Astrophys. Space. Sci} \textbf{343}, 559 (2013).
\bibitem{CSKKGBH}J. sadeghi, and H. Vaez, arXiv:1310.4486 [gr-qc].

\bibitem{saadat}H. Saadat, \emph{Int J theor Phys}, [arXiv:1306.0601v1
[gr-qc]].

\bibitem{NarasimhaChitre}D. Narasimha and S. M. Chitre, \emph{Astro. J}. \textbf{97} 327
(1989).
\bibitem{Narasimha}D. Narasimha, \emph{ICNAPP}, ed. R. Cowsik,
\textbf{251} (1994).
\bibitem{hogan}C. J. Hogan, R. Narayan, \emph{MNRAS} \textbf{2111}, 575 (1984).
\bibitem{Vilenkin1}A. Vilenkin, \emph{ApJL} \textbf{51}, 282 (1984).
\bibitem{Vilenkin2}A. Vilenkin, \emph{Nat} \textbf{322}, 613 (1986).




\bibitem{Zwiebach}B. Zwiebach, \emph{Phys. Lett. B} 156, 315 (1985).

\bibitem{zumino}B. Zumino, \emph{ Phys. Rep}.\textbf{ 137}, 109 (1986).




\bibitem{Boulware}D. G. Boulware and S. Deser, \emph{Phys. Rev. Lett}. \textbf{55},
2656 (1985).

\bibitem{wiltshire}D. Wiltshire, \emph{Phys. Rev. D} \textbf{38}, 2445 (1988).
\bibitem{Rong1}Rong-Gen Cai ei, Qi Guo, \emph{Phys. Rev. D} \textbf{69}, 104025 (2004).
 \bibitem{Rong2}Rong-Gen Cai, \emph{Phys. Rev. D} \textbf{65}, 084014 (2002).
\bibitem{}J. Crisostomo, R. Troncoso, J. Zanelli, \emph{Phys. Rev. D} \textbf{62}, 084013 (2000).
\bibitem{}Sachiko Ogushi, Misao Sasaki, \emph{Prog. Theor. Phys}. \textbf{113} 979
(2005).
\bibitem{}Tim Clunan, Simon F. Ross, Douglas J. Smith, \emph{Class. Quant. Grav}.
\textbf{21},
3447 (2004).
\bibitem{}A. Barrau, J. Grain, S.O. Alexeyev, \emph{Phys.Lett.B} \textbf{584}, 114
(2004).
\bibitem{}Maximo Banados, \emph{Phys. Lett. B} \textbf{579}, 13
(2004).
\bibitem{}M. Cvetic, S. Nojiri, S. D. Odintsov, \emph{Nucl. Phys. B} \textbf{ 628}, 295
(2002).
\bibitem{}S. Nojiri, S. D. Odintsov, \emph{Phys. Lett. B} \textbf{523}, 165 (2001).
\bibitem{}S. Nojiri, S. D. Odintsov, S. Ogushi, \emph{Phys. Lett. D} \textbf{65}, 023521
(2002).
\bibitem{}Gustavo Dotti, Reinaldo J. Gleiser, \emph{Class. Quant. Grav.} \textbf{22}, L1
(2005).
\bibitem{}Ishwaree P. Neupane, \emph{Phys. Lett. D} \textbf{69}, 084011
(2004).
\bibitem{Ishwaree}Ishwaree P. Neupane, \emph{Phys. Lett. D} \textbf{67}, 061501
(2003).
\bibitem{Gillessen} S. Gillessen, F. Eisenhauer, S. Trippe, T. Alexander, R. Genzel, F. Martins, and T. Ott, \emph{ApJ} \textbf{692}, 1075
(2009).

\end{thebibliography}
\end{document}